\documentclass[aps,pra,preprint,groupedaddress,showpacs,eqsecnum]{revtex4}

\usepackage[dvips]{graphicx}

\begin{document}

\title{Generation of entangled states of two atoms inside a leaky cavity}
\author{T.~W.~Chen, C.~K.~Law, and P.~T.~Leung}
\affiliation{Department of Physics, The Chinese University of Hong Kong, \\
Shatin, Hong Kong SAR, China}
\date{\today }

\begin{abstract}
An in-depth theoretical study is carried out to examine the
quasi-deterministic entanglement of two atoms inside a leaky
cavity. Two $\Lambda$-type three-level atoms, initially in their
ground states, may become maximally entangled through the
interaction with a single photon. By working out an exact analytic
solution, we show that the probability of success depends
crucially on the spectral function of the injected photon. With a
cavity photon, one can generate a maximally entangled state with a
certain probability that is always less than $50\%$. However, for
an injected photon with a narrower spectral width, this
probability can be significantly increased. In particular, we
discover situations in which entanglement can be achieved in a
single trial with an almost unit probability.
\end{abstract}

\maketitle

\section{Introduction} \label{sec:Introduction}
Entanglement has long been considered a fundamental property in
quantum theory~\cite{Schrodinger-Entanglement,EPR,Bell}. It has
gained renewed interest due to its potential applications in
quantum computations and information in recent years, and is
generally considered as the corner-stone of this field (see, e.g.,
Refs.~\cite{Bo-book,Chuang-book,Haroche-entanglement-RMP} and
references therein). While entanglement can be generated by a lot
of different implements such as solid-state
devices~\cite{Bouchiat,Bertoni,Friedman,Lloyd-sci-2}, nuclear
magnetic resonance in liquid
samples~\cite{Chuang-sci,Jones-nature}, and ion
trap~\cite{Monroe-gate,Turchette-entangle-trap,Sackett-nature}, it
is most conveniently realized by photons, e.g., in parametric
down-conversion~\cite{Yamamoto-nature,two-photon-imaging}.
However, though being ideal carriers of quantum information,
photons cannot serve as convenient storage media. The storage of
information is much better implemented with atoms. In this sense,
cavity quantum electrodynamics (CQED) provides a perfect stage for
the interaction of information carriers and depository to take
place~\cite{Kimble-cesium,Haroche-phase-gate,Memory-photon}. With
the advance in CQED technologies, some proposals have also been
made to generate entangled states of atoms in optical and
microwave cavities~\cite{Plenio,
Beige,Pachos,Probabilistic-entangle-cavity,Haroche-EPR,Duan-multiatom-entangle},
many of which employ the leakage of the cavity not only as a
channel for information passage, but also as the crucial element
for generating entanglement~\cite{Braun,Plenio}. In addition,
state projection techniques using quantum measurement have also
been proposed to generate entanglement in a probabilistic manner
\cite{Probabilistic-entangle-cavity,Duan-multiatom-entangle}.

Recently, Hong and Lee have proposed a scheme to generate
entangled atoms inside a leaky cavity in a ``quasi-deterministic"
manner~\cite{Hong}. In their model, which is similar to that in
Ref.~\cite{Plenio}, two $\Lambda$-type three-level atoms,
initially prepared in their ground states, may become maximally
entangled after interacting with a single cavity photon. The
polarization of the output photon, serving as an indicator of the
atomic state, is then measured, and generation of entangled atomic
state is successful if the photon carries the right polarization.
Adopting the master equation approach, they numerically showed
that under the optimal condition in which the ratio of the
coupling strengths of the two channels is $\sqrt{2}$, the maximum
probability of success in each operation is $50\%$. They further
argued that a unit probability of success can be achieved by
introducing the output photon to the leaky cavity repeatedly.
Through the application of this automatic feedback scheme, the two
atoms in the cavity are entangled in a ``quasi-deterministic"
manner.

The method proposed by Hong and Lee is appealing. However, their
scheme is still probabilistic for each single trial. It is
therefore interesting to study whether the probability of success
in each trial could be increased.  The master equation approach
adopted in Ref.~\cite{Hong} has limited further development in
this direction. In addition, they obtained the optimal condition
and hence the maximum probability by direct inspection on the
numerical results. In the present paper, we first develop a
rigorous analytic solution to their model and in turn provide a
physical interpretation for the optimal condition mentioned above.
We are then able to show explicitly  that the upper bound $50\%$
of the probability of success is an intrinsic nature of their
scheme that makes use of a cavity photon.

Furthermore, by generalizing the initial photon state to consider
photons injected into the cavity with certain spectra, we find
that the probability of success in each trial can be increased and
larger than $50\%$. It is worthwhile to note that the master
equation approach employed by Hong and Lee in Ref.~\cite{Hong} is
not a proper description for the injected and output photons,
which are usually characterized by wave packets with certain
durations or, equivalently, by their respective spectral
functions. The master equation approach, on the other hand, deals
mainly with the ``decay" of photons existing (or created) in the
cavity. A photon that has already leaked out of the cavity is
considered as a ``decay" event and is no longer part of the
system. As a consequence, their approach fails to describe the
quantum states of the injected and output photons and is therefore
deemed inappropriate in this situation.

In this paper, we adopt the continuous frequency mode approach
that considers the cavity and its environment as a single entity
to quantize the entire
system~\cite{Loudon,Lai-narrow-resonance,Scully-MOU,JC-pure-CK}.
This enables us to study a photon with an arbitrary spectral
function. We find that the probability of success in each
operation depends crucially on the spectral function of the
injected photon. In general, the probability can be increased to
well over $50\%$. As long as spontaneous decay can be neglected,
with suitably chosen spectral function of the photon,
deterministic entanglement generation can be realized irrespective
of the leakage rate of the cavity. Remarkably, an almost unit
probability of success can be obtained for realistic CQED setups
and photon packets.

The structure of our paper is as follows: In Sec.~\ref{sec:MOU},
we introduce a proper set of spatial mode functions --- the
continuous frequency modes, in terms of which the electromagnetic
field is quantized. The interaction Hamiltonian and the evolution
of the atoms governed by it are discussed in
Sec.~\ref{sec:atom-field interaction} and \ref{sec:Calculation},
respectively. In Sec.~\ref{sec:Generation-entanglement}, the
generation of entanglement is studied for two cases with different
input states, namely, a cavity
photon~\cite{Lang_PRA,JC-pure-CK,plank} and an arbitrary injected
photon. We conclude the main features of our scheme in
Sec.~\ref{sec:Conclusion}.

\section{Continuous frequency modes}\label{sec:MOU}
\begin{figure}
\includegraphics[width=8.6cm]{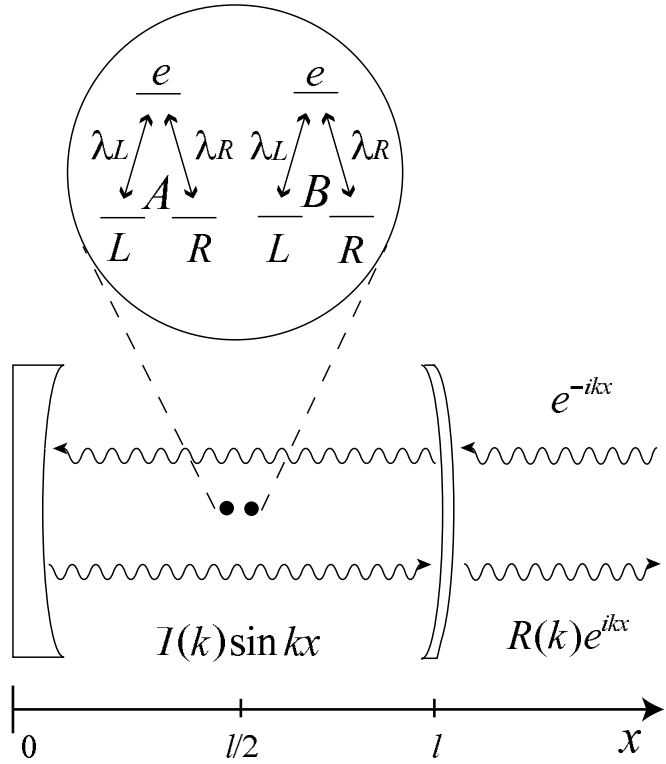}
\caption{A sketch of the system in our model. Two $\Lambda$-type
three-level atoms are trapped near the center of a one-sided leaky
cavity. The continuous frequency modes shown in the figure are
adopted to study the generation of atomic entangled state by a
cavity mode photon or an injected photon packet.} \label{fig1}
\end{figure}

In the following discussion we adopt the continuous frequency mode
quantization scheme to analyze  optical processes taking place
inside a leaky optical
cavity~\cite{Loudon,Lai-narrow-resonance,Scully-MOU,JC-pure-CK}.
Hence photons are represented by quantum states $|k_{L}\rangle $
and $|k_{R}\rangle $, which are characterized by a continuous wave
number $k$, and the polarization ($L$ and $R $) of the photon. For
definiteness and without loss of generality, we consider in this
paper a one-sided optical cavity as shown in Fig.~\ref{fig1},
which has length $l$ and is bounded by two mirrors. The left
mirror at $x=0$ is perfectly reflecting; the right mirror at $x=l$
is partially transparent and characterized by frequency-dependent
reflection and transmission coefficients $r(k)$ and $t(k)$. The
spatial mode functions $U_k(x)$ corresponding to quantum states
$|k_{L(R)}\rangle $ are given by
\begin{equation}
\label{Uk}
U_k(x)=\left\{\begin{array}{c} I(k)\sin{kx}\\
e^{-ikx}+R(k)e^{ikx} \end{array} \right.
\begin{array}{ccc}{}&{}&{}\\
{}&{}&{}
\end{array}
\begin{array}{c}
0<x\leq l,\\
l<x<\infty,
\end{array}
\end{equation}
where
\begin{eqnarray}
\label{Ik}
I(k)&=&\frac{-2it}{1+re^{2ikl}}, \\
\label{Rk} R(k) &=& \frac{-r-t+re^{-2ikl}}{1+re^{2ikl}}.
\end{eqnarray}
Notice that $|R(k)|=1$, as required by conservation of energy.
Standard Sturm-Liouville type orthogonality integral leads to the
general normalization condition
\begin{equation}
\int_0^\infty
\epsilon(x)U_k(x)U^\ast_{k^{\prime}}(x)dx=2\pi\delta(k-k^{\prime}),
\end{equation}
where $\epsilon(x)$ is the position-dependent relative
permittivity. For example, if an infinitely thin dielectric mirror
is placed at $x=l$, then $\epsilon(x)=1+\zeta \delta(x-l)$, where
$\zeta$ determines the finesse of the mirror~\cite{Loudon}.

The quantization of the field is hence accomplished by decomposing
the field operator $\hat{A}(x,t)$ into these continuous frequency
modes and by introducing the usual annihilation (creation)
operator $\hat{a}_{k\mu}$ ($\hat{a}^\dag_{k\mu}$), with $\mu=L,R$
being the polarization index. These operators satisfy the standard
commutation relation $[\hat{a}_{k\mu},
\hat{a}^\dag_{k'\mu'}]=\delta_{\mu\mu'}\delta(k-k^{\prime})$. The
result reads:
\begin{equation}
\label{field-operator} \hat{A}(x,t)=\sum\limits_{\mu=L,R}
\int_0^\infty
k^{-1/2}\left[U_k(x)\hat{a}_{k\mu}e^{-ikt}+U^\ast_k(x)\hat{a}^\dag_{k\mu}e^{ikt}\right]dk
.
\end{equation}
Hereafter we adopt the cgs units and take $\hbar=c=1$.

\begin{figure}
\includegraphics[width=8.6cm]{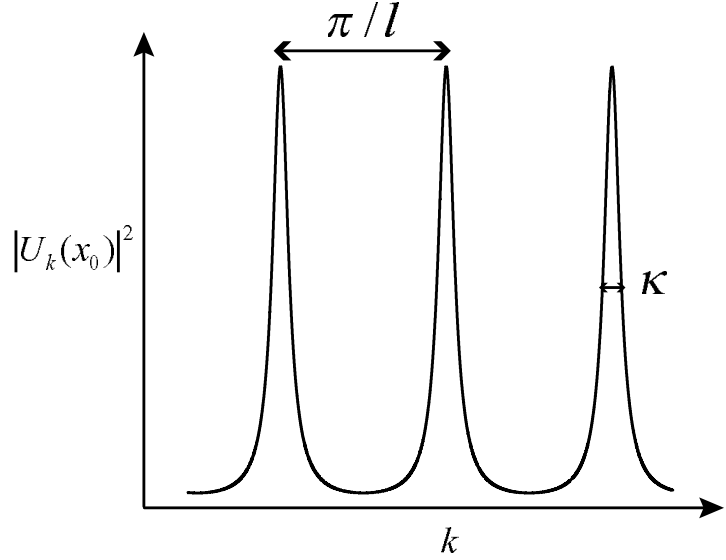}
\caption{A schematic plot of $|U_k(x_0)|^2$ versus the frequency
$k$ at a point $x_0$ inside the cavity. Each peak corresponds to a
quasi-mode of the cavity. The separations and widths of the
quasi-modes are given by $\pi/l$ and $\kappa$, respectively. In
our model, we assume that the atomic frequency is close to one of
the quasi-mode frequencies, and adopt the single mode
approximation.} \label{fig2}
\end{figure}

The coefficients $I(k)$ and $R(k)$ possess poles $\tilde{k}_n$,
which are roots of the equation $1+re^{2ikl}=0$. As long as $r(k)$
and $t(k)$ are slowly varying functions of $k$, the roots of the
above equation are approximately given by: $\tilde{k}_n=k_n
-i\kappa_n/2$, where
\begin{eqnarray}
\label{resonance-frequencies} k_n&=&\frac{(2n+1)\pi-\arg
\left[r(k_{n})\right]}{2l}
, \\
\label{kappa} \kappa_n&=&-\frac{\log{|r(k_{n})|}}{l} ,
\end{eqnarray}
and $n=0,\pm 1,\pm2,\ldots$ . These complex frequencies define the
quasi-modes of the
cavity~\cite{Loudon,Lai-narrow-resonance,Scully-MOU,QNM-RMP}.
Figure~\ref{fig2} shows a sketch of $|U_k (x_0)|^2$ versus $k$ at
a point $x_0$ inside the cavity. Each peak in the figure
corresponds to a quasi-mode of the cavity. For large $n$, the
separations and widths of the modes can be taken as constants
given by $\pi/l$ and $\kappa$, respectively. The physical
significance of these quasi-modes is that for a ``good cavity",
where $|r|\simeq 1$, the mode function is negligibly small except
at frequencies $k \simeq k_n$. In fact, it can be shown that the
electrodynamics inside the cavity is completely describable in
terms of these quasi-modes~\cite{QNM-RMP}. In the following, we
shall make use of this formalism to discuss the interaction
between the atoms and the quantized photon field.

\section{Atom-field interaction} \label{sec:atom-field interaction}
We consider two identical $\Lambda$-type atoms $A$ and $B$ located
near the center of a one-sided cavity as shown in Fig.~\ref{fig1}.
The separation of the atoms is assumed to be much smaller than the
wavelength of the field in interest. The two ground states and the
excited state are, respectively, denoted by $|L_{\alpha}\rangle $,
$|R_{\alpha}\rangle $ and $|e_{\alpha }\rangle $, where $\alpha
=A$, $B$ refers to atom $A$ and $B$. The excited state couples
with the ground state $|L_{\alpha}\rangle $ ($|R_{\alpha}\rangle
$) by emitting a $|k_{L}\rangle $ ($|k_{R}\rangle $) mode photon.
In our model, we assume that the ground states are stable, and
the spontaneous decay rate of the excited states into non-cavity
modes is negligible compared with the cavity leakage rate
$\kappa$. We shall use the notation $|\mu_A\mu_B;k_{\mu}\rangle$
for states with two ground-state atoms and a single photon, where
$\mu_A$, $\mu_B$, $\mu=L$, $R$ and the first and second positions
are assigned to atom $A$ and $B$, respectively. States with an
excited-state and a ground-state atom will similarly be denoted
by $|e_A\mu_B;0\rangle$ and $|\mu_A e_B;0\rangle$, where
$|0\rangle$ is the vacuum state of the field.

Assuming that the two ground states are degenerate and the
energies of the ground and excited states are zero and $\omega_e$,
respectively, in terms of the continuous frequency mode basis, the
Hamiltonian can be written as
\begin{eqnarray}
\nonumber
\hat{H}&=&\sum\limits_{\alpha=A,B}\omega_{e}|e_{\alpha }\rangle
\langle e_{\alpha }|+\int_0^{\infty} dk
\sum\limits_{\mu=L,R}\omega_{k}a_{k\mu }^{\dag }a_{k\mu } \\
\label{H}
&&+\int_0^{\infty} dk \sum\limits_{\alpha=A,B \atop \mu=L,R}g_{\mu
}(k)a_{k\mu }|e_\alpha\rangle \langle
\mu_{\alpha}|+\mathrm{h}.\mathrm{c}. ,
\end{eqnarray}
where the function $g(k)$ is proportional to the mode function
$U_k(x)$ evaluated at the position of the atoms, $x_a \simeq l/2$.

In the present paper, the atomic frequency is assumed to be close
to one of the cavity quasi-mode frequencies given by
Eq.~(\ref{resonance-frequencies}). Therefore, one can expand the
mode functions around the resonance frequency and obtain the
so-called single mode approximation result
\cite{JC-pure-CK,Lang_PRA,plank}:
\begin{equation}
g_{\mu }(k)=\frac{\sqrt{\kappa/2\pi}\lambda_{\mu }e^{i\theta _{\mu
}}}{k-k_{c}+i\kappa/2} ,  \label{gk}
\end{equation}
where $k_c$ and $\kappa$ are, respectively, the frequency and the
decay rate of the quasi-mode in resonance, and $\theta _{\mu }$
is a trivial phase angle. Besides,
\begin{equation}
\lambda_{\mu}^{2}=\int_{-\infty}^{\infty}|g_{\mu }(k)|^{2}dk\
\label{g-square}
\end{equation}
is a measure of the coupling strength that depends on the dipole
moment and the location of the atom. Also notice that within the
single mode approximation, the lower limit of the frequency domain
will hereafter be extended from $0$ to $-\infty$, as shown in
Eq.~(\ref{g-square}).

\section{Analytic solution} \label{sec:Calculation}
The system is initially prepared in the state
\begin{equation}
\label{initial-state} |\Psi \rangle _{\rm
in}=\int_{-\infty}^{\infty} dk^\prime
f(k^\prime)|LL;k^\prime_L\rangle ,
\end{equation}
where $f(k^\prime)$ is the spectral function of a one-photon state
in the continuous frequency mode representation, satisfying the
normalization condition:
\begin{equation}
\int_{-\infty}^{\infty} |f(k^\prime)|^{2}dk^\prime=1 .
\label{normalization}
\end{equation}
The explicit form of the injected photon will be specified later.

We introduce an essential state basis to simplify the dynamics of
the two atoms as follows. The frequency and polarization of the
photon are two independent degrees of freedom of the Hilbert
space. Restricted to single excitation, one of the two atoms may
be excited while the other can be in $|L\rangle$ or $|R\rangle$,
and thus there are four distinct states with a single excited
atom. When the two atoms are in the ground states ($|L\rangle$ and
$|R\rangle$ states), the polarization of the emitted photon can be
either $L$ or $R$, giving rise to a total of eight distinct states
for each frequency. However, if we are only interested in the
evolution with the specific initial state given by
Eq.~(\ref{initial-state}), the dimension of the subspace involved
is greatly reduced. One of the atoms in the initial state
$|LL;k^\prime_L\rangle$ may absorb the photon and the state
evolves into
\begin{equation}
\label{E}
|E\rangle=\frac{1}{\sqrt{2}}\left(|eL;0\rangle+|Le;0\rangle\right)
.
\end{equation}
When the excited atom de-excites by emitting another photon of
wave number $k$ ($k \ne k^{\prime}$ in general), the resulting
state may be $|LL;k_L\rangle$ or $|\Phi ;k_R\rangle$, where
\begin{equation}
\label{entangle}
|\Phi\rangle=\frac{1}{\sqrt{2}}\left(|LR\rangle+|RL\rangle\right)
\end{equation}
is a Bell state of the two atoms --- the goal of the current
discussion. Hence, the dynamics of our system is adequately
described by transitions between the excited state $|E\rangle$ and
the single-photon states $|LL;k_L\rangle$ or $|\Phi;k_R\rangle$.
Considering hereafter an initial state specified by
Eq.~(\ref{initial-state}), we obtain an effective Hamiltonian
within the above-mentioned subspace:
\begin{eqnarray}
\nonumber
\hat{H}_{\rm eff} &=& \omega_{e}|E\rangle
\langle E|+\int_{-\infty}^{\infty} dk \, k
\left(|LL;k_L\rangle\langle
LL;k_L|+|\Phi;k_R\rangle\langle\Phi;k_R|\right) \\
&&+\int_{-\infty}^{\infty} dk
\left(\sqrt{2}g_L(k)|E\rangle\langle
LL;k_L|+g_R(k)|E\rangle\langle\Phi;k_R|\right)
+\mathrm{h}.\mathrm{c}. .
\end{eqnarray}

It is noteworthy that the dynamics of the current system, governed
by $\hat{H}_{\rm eff}$, bears strong resemblance to that of a
single $\Lambda$-type three-level system in an optical cavity,
with coupling strengths of the two channels being in the ratio of
$\sqrt{2}\lambda_L$ to $\lambda_R$. The factor $\sqrt{2}$ can be
explained as follows. First, while both atoms in state
$|LL;k_L\rangle$ can interact with the photon, one of the two
atoms in state $|\Phi;k_R\rangle$ is only a spectator. Second, the
collective quantum effect of the state $|\Phi\rangle$ contributes
an enhancement factor $\sqrt{2}$. These two effects together give
an overall $\sqrt{2}$ factor. Hence, if the coupling strengths
$\lambda_L$ and $\lambda_R$ are in the ratio of $1:\sqrt{2}$,
satisfying the optimal condition discovered in Ref.~\cite{Hong},
the original two-atom model reduces to a symmetric three-level
system. Interestingly enough, it is well known that a resonantly
driven symmetric $\Lambda$-type atom located in an ideal cavity
can exhibit perfect oscillations between its two ground states. We
shall see that these analogies do shed light on our current
investigation.

To simplify the calculations, we perform another basis
transformation to reduce the model to a two-level system. By
defining
\begin{eqnarray}
\label{psi1} |\psi_1(k)\rangle &=&
\frac{1}{V(k)}\left(\sqrt{2}g_L^{\ast}(k)|LL;k_L\rangle+g_R^{\ast}(k)|\Phi;k_R\rangle\right)
, \\
\label{psi2} |\psi_2(k)\rangle &=&
\frac{1}{V(k)}\left(g_R(k)|LL;k_L\rangle-\sqrt{2}g_L(k)|\Phi;k_R\rangle\right)
,
\end{eqnarray}
where
\begin{equation}
\label{Vk} V(k)=\sqrt{2|g_L(k)|^2+|g_R(k)|^2},
\end{equation}
one of the ground states is turned into a dark state
$|\psi_2(k)\rangle$, which does not take part in the interaction.
The Hamiltonian can thus be written as
\begin{equation}
\hat{H}_{\rm eff}=\hat{H}_0+\hat{V}+\hat{H}_{\rm dark},
\end{equation}
where
\begin{eqnarray}
\label{H0-eff-psi}
\hat{H}_{0}&=&\omega_{e}|E\rangle
\langle E|+\int_{-\infty}^{\infty} dk \, k
|\psi_1(k)\rangle\langle\psi_1(k)|, \\
\label{V-eff-psi}
\hat{V}&=&\int_{-\infty}^{\infty} dk \,
V(k)|E\rangle\langle\psi_1(k)|+\mathrm{h}.\mathrm{c}., \\
\label{Hdark}
\hat{H}_{\rm dark}&=&\int_{-\infty}^{\infty}dk \, k
|\psi_2(k)\rangle\langle\psi_2(k)|,
\end{eqnarray}
with $\hat{H}_{\rm dark}$ being the free Hamiltonian of the dark states
$|\psi_2(k)\rangle$.

To study the output state $|\Psi\rangle_{\rm out}$ of the system
subject to an arbitrary state $|\Psi\rangle_{\rm in}$ at $t=0$, we
adopt the resolvent method (see, e.g Ref.~\cite{Cohen}) to deal
with transitions between states with different $k$'s.
Remarkably, if the trivial evolution of the dark state is
discarded, the dynamics governed by the effective Hamiltonian is
ostensibly analogous to that of a two-level atom interacting with
a single quasi-mode. To simplify the calculation, the dark states
will be ignored when solving the system using resolvent method.

The resolvent of the Hamiltonian $\hat{H}_0+\hat{V}$ is given by
\begin{equation}
\hat{G}(\omega)=\frac{1}{\omega-\hat{H}_0-\hat{V}} ,  \label{Gw}
\end{equation}
with $\omega$ being a complex variable. It yields the retarded
Green's function, $\theta(t)\exp[-i(\hat{H}_0+\hat{V}) t]$,
through an integral transformation:
\begin{equation}
\theta(t)\exp[-i(\hat{H}_0+\hat{V}) t]
=\lim_{\epsilon\rightarrow0^+}\frac{i}{2\pi}\int_{-\infty+i\epsilon}^{\infty+i\epsilon}
\hat{G}(\omega)e^{-i\omega t}d\omega. \label{Evo-Gw}
\end{equation}
In the following discussion, unless otherwise stated, we assume
$t>0$ and therefore identify the retarded Green's function with
the evolution operator
$\hat{U}(t,0)\equiv\exp[-i(\hat{H}_0+\hat{V})  t]$.

To evaluate the matrix elements of the resolvent, we partition the
Hilbert space into two complementary parts with projection
operators $\hat{P}=|E\rangle \langle E|$ and $\hat{Q}={\bf
1}-\hat{P}$. The projection of $\hat{G}(\omega)$ by $\hat{P}$ is
given by~\cite{Cohen}:
\begin{equation}
\hat{P}\hat{G}(\omega)\hat{P}=\frac{\hat{P}}{\omega-\hat{P}\hat{H}_{0}\hat{P}-\hat{P}\hat{R}(\omega)\hat{P}}
, \label{PGwP}
\end{equation}
where $\hat{R}(\omega)$ is the level-shift operator defined as
\begin{equation}
\hat{R}(\omega)=\hat{V}+\hat{V}\frac{\hat{Q}}{\omega-\hat{Q}\hat{H}_{0}\hat{Q}-\hat{Q}\hat{V}\hat{Q}}\hat{V}
. \label{Rw}
\end{equation}
It is then straightforward to show that
\begin{eqnarray}
\langle E|\hat{R}(\omega)|E\rangle &=&\int_{-\infty}^{\infty} \frac{V(k)^2}{\omega-k} dk  \nonumber \\
&=&\frac{2
\lambda_{L}^{2}+\lambda_{R}^{2}}{\omega-k_{c}+i\kappa/2} .
\label{REE}
\end{eqnarray}
Hence,
\begin{eqnarray}
\label{GEE1} \langle E| \hat{G}(\omega)|E\rangle &=&\frac{\Delta
\omega+i\kappa/2}{(\Delta \omega-\delta_e)(\Delta
\omega+i\kappa/2)-2
\lambda_L^2-\lambda_R^2} \\
\label{GEE2} &=&\frac{\Delta \omega +i\kappa/2}{(\Delta
\omega-\omega_+)(\Delta \omega-\omega_-)} ,
\end{eqnarray}
with
\begin{eqnarray}
\label{Dw}
\Delta \omega&=&\omega-k_c , \\
\label{detune} \delta_e &=& \omega_e-k_c ,
\end{eqnarray}
and
\begin{equation}
\omega_{\pm}=\frac{\delta_e-i\kappa/2}{2}\pm\sqrt{\left(\frac{\delta_e+i\kappa/2}{2}\right)^2+2\lambda_L^2+\lambda_R^2}.
\end{equation}
In our model, the atoms are in ground state at the beginning and
the end. Hence the above amplitude is not of direct interest.
However, this result is essential for the evaluation of the
ground-state amplitudes, which is our major objective. The
projection of the resolvent by $\hat{Q}$ is
\begin{eqnarray}
\hat{Q}\hat{G}(\omega)\hat{Q} &=&\frac{\hat{Q}}{\omega-\hat{Q}\hat{H}_{0}\hat{Q}-\hat{Q}\hat{V}\hat{Q}}  \nonumber \\
&&+\frac{\hat{Q}}{\omega-\hat{Q}\hat{H}_{0}\hat{Q}-\hat{Q}\hat{V}\hat{Q}}\hat{V}\frac{\hat{P}}{\omega-\hat{P}\hat{H}_{0}\hat{P}-\hat{P}\hat{R}(\omega)\hat{P}}
\nonumber \\
&&
\hat{V}\frac{\hat{Q}}{\omega-\hat{Q}\hat{H}_{0}\hat{Q}-\hat{Q}\hat{V}\hat{Q}}
, \label{QGwQ}
\end{eqnarray}
yielding the ground-state matrix elements
\begin{equation}
\label{Gkk'} \langle
\psi_1(k)|\hat{G}(\omega)|\psi_1(k^{\prime})\rangle =
\frac{\delta(k-k^{\prime})}{\omega-k}+\frac{V(k)V(k^{\prime})}{(\omega-k)(\omega-k^{\prime})}
\langle E|\hat{G}(\omega)|E\rangle .
\end{equation}
These  expressions are generally
valid~\cite{Resolvent-continuation}, and can be applied to any
initial photon states. In addition, when we are interested in an
initial state $|\Psi\rangle_{\rm in}$ satisfying the scattering
state condition:
\begin{equation}
\label{scatter-cond}
\begin{array}{ccccc}
\hat{V}e^{-i\hat{H}_0t}|\Psi\rangle_{\rm in}=0 &&& {\rm for }&
t\le0,
\end{array}
\end{equation}
then it can be shown that:
\begin{eqnarray}
\label{scatter-1}
\langle\psi_1(k)|\hat{U}(t,0)|\psi_1(k^{\prime})\rangle
&=&\delta(k-k^{\prime})e^{-ikt}\left[1-2\pi i V(k)^{2}\langle
E|\hat{G}(k)|E\rangle\right] ,\\
\label{scatter-2} &=& \delta(k-k^{\prime})e^{-ikt} e^{i
\delta_{\rm s}(k)},
\end{eqnarray}
with $\delta_{\rm s}(k)$ being a $k$-dependent real phase shift.
Equation~(\ref{scatter-2}) is obtained by noticing that the module
of the last factor in Eq.~(\ref{scatter-1}) is one. While details
of the proof will be given in the appendix, the physical meaning
of Eq.~(\ref{scatter-cond}) is obvious. It simply implies that the
atoms do not experience the field of the injected photon as long
as $t\leq 0$. Equations~(\ref{scatter-1}) and (\ref{scatter-2})
become useful when we consider an injected photon prepared in a
scattering state.

\section{Generation of entanglement} \label{sec:Generation-entanglement}
After interacting with the atoms, the photon leaks out of the
cavity, and its polarization can be detected, e.g. by a quarter
wave plate and a polarization beam splitter. The atoms inside the
cavity will then be projected into either the direct product state
$|LL\rangle$ or the maximally entangled state $\Phi=(|LR\rangle
+|RL\rangle )/\sqrt{2}$, depending, respectively, on whether an
$L$ or $R$ mode photon is detected. In the following, we consider
two different cases where the photon in the initial state is (i)
prepared in a cavity mode; and (ii) injected from outside with a
spectral function satisfying Eq.~(\ref{scatter-cond}).

\subsection{Cavity photon} \label{sub:Cavity-photon}
A cavity photon is identified with a quasi-mode photon as defined
in Ref.~\cite{plank}, which is initially created inside the cavity
and leaks gradually to the surroundings. In mathematical terms, a
cavity photon is characterized by the spectral function:
\begin{equation}
f(k^\prime)=f_c(k^\prime)=\frac{1}{\lambda_{\mu }} g_{\mu }^{\ast
}(k^\prime), \label{cavity-mode}
\end{equation}
which corresponds to the cavity line shape.
Equations~(\ref{cavity-mode}), (\ref{psi1}), and (\ref{psi2}) then
lead directly to
\begin{equation}
|\Psi\rangle_{\rm in}=\int_{-\infty}^{\infty}
dk^{\prime}\frac{f_c(k^{\prime})}{V(k^{\prime})}\left(\sqrt{2}g_L(k^{\prime})|\psi_1(k^{\prime})\rangle
+g_R^{\ast}(k^{\prime})|\psi_2(k^{\prime})\right) .
\end{equation}
The evolution of the dark states $|\psi_2(k^\prime)\rangle$ is
trivial, whereas the resolvent method gives the Fourier transform
of the time evolution of the states $|\psi_1(k^\prime)\rangle$. As
photon detections are carried out at the end, the relevant matrix
elements of the resolvent are those given by Eq.~(\ref{Gkk'}).
Hence,
\begin{equation}
\langle \psi_1(k)|\hat{G}(\omega)|\Psi\rangle_{\rm
in}=\frac{\sqrt{2}}{\omega-k}\left(\frac{g_L(k)f_c(k)}{V(k)}+\frac{\lambda_L
\langle E|\hat{G}(\omega)|E\rangle
V(k)}{\omega-k_c+i\kappa/2}\right) ,
\end{equation}
and from inverse Fourier transform we obtain, as $t \rightarrow
\infty$,
\begin{equation}
\langle \psi_1(k)|\hat{U}(t,0)|\Psi\rangle_{\rm
in}=\sqrt{2}\left(\frac{g_L(k)f_c(k)}{V(k)} +\frac{\lambda_L
V(k)}{(\Delta k-\omega_+)(\Delta k-\omega_-)}\right)e^{-ikt} ,
\end{equation}
where all the transients have been neglected, and
\begin{equation}
\Delta k=k-k_c .
\end{equation}
Notice that although we are looking at the long time behavior,
Eqs.~(\ref{scatter-1}) and (\ref{scatter-2}) do not apply to this
situation because the initial state given by $f_c(k^\prime)$ does
not satisfy Eq.~(\ref{scatter-cond}).

The resolvent method only solves the dynamics of the states
$|\psi_1(k)\rangle$.
The overall output state also includes the contributions of the dark
states, in which a trivial phase factor $e^{-ikt}$ is multiplied. With the
dark state included again, the output state in the long time limit is
given by \begin{eqnarray}
|\Psi \rangle _{\rm out} &=&|LL\rangle \otimes
\int_{-\infty}^{\infty} dk \, e^{-ikt}f_c(k)\frac{(\Delta
k-\delta_e)(\Delta k+i\kappa/2)-\lambda_R^2}
{(\Delta k-\omega_+)(\Delta k-\omega_-)} |k_{L}\rangle   \nonumber \\
&&+|\Phi \rangle \otimes \int_{-\infty}^{\infty} dk \,
e^{-ikt}f_c(k)\frac{\sqrt{2} \lambda_{L}\lambda_{R}e^{i\Delta
_{LR}}}{(\Delta k-\omega_+)(\Delta k-\omega_-)}|k_{R}\rangle ,
\label{psiout-cavity}
\end{eqnarray}
where
\begin{equation}
\Delta _{LR}=\theta_{L}-\theta _{R} .  \label{DeltaLR}
\end{equation}

If one then detects the photon in the $L$ ($R$) mode, the atomic
state is projected into the non-entangled state $|LL\rangle$ (the
maximally entangled state $|\Phi \rangle$), with respective
probabilities $P_{L}$ and $P_{R}$ given by:
\begin{eqnarray}
P_{L} &=&\int_{-\infty}^{\infty} \left| f_c(k)\frac{(\Delta
k-\delta_e)(\Delta k+i\kappa/2)-\lambda_R^2}{(\Delta
k-\omega_+)(\Delta k-\omega_-)} \right| ^{2} dk
\nonumber \\
&=&1-\frac{4\lambda_{L}^{2}\lambda_{R}^{2}(2\lambda_{L}^{2}+
\lambda_{R}^{2}+\kappa^{2}/2)}{(2\lambda_{L}^{2}+\lambda
_{R}^{2})\left[ (2\lambda_{L}^{2}+\lambda_{R}^{2}+\kappa^{2}/2)^{2}+\delta_{e}^{2}\kappa^{2}\right] } ,  \label{PL} \\
P_{R} &=&2\lambda_{L}^{2}\lambda_{R}^{2}\int_{-\infty}^{\infty} \left| \frac{f_c(k)}{(\Delta k-\omega_+)(\Delta k-\omega_-)}\right| ^{2} dk  \nonumber \\
&=&\frac{4\lambda_{L}^{2}\lambda_{R}^{2}(2\lambda_{L}^{2}+
\lambda_{R}^{2}+\kappa^{2}/2)}{(2\lambda_{L}^{2}+\lambda
_{R}^{2})\left[
(2\lambda_{L}^{2}+\lambda_{R}^{2}+\kappa^{2}/2)^{2}+\delta_{e}
^{2}\kappa^{2}\right] } . \label{PR}
\end{eqnarray}
If the resonance condition is satisfied, i.e., $\delta_e =0$, we
have
\begin{equation}
P_{R}=\frac{4\lambda_{L}^{2}\lambda_{R}^{2}}{(2\lambda
_{L}^{2}+\lambda_{R}^{2})(2\lambda_{L}^{2}+\lambda
_{R}^{2}+\kappa^{2}/2)} .  \label{P-no detuning}
\end{equation}
In strong coupling regime, where $\lambda_{L}$, $\lambda_{R}\gg
\kappa$,
\begin{equation}
P_{R}\cong \left[ \frac{2\lambda_{L}\lambda_{R}}{(2\lambda
_{L}^{2}+\lambda_{R}^{2})}\right] ^{2} .  \label{PR-strong
coupling}
\end{equation}
One can easily prove that
\begin{equation}
P_{R}\leq \frac{1}{2}  \label{PRmax-strong coupling}
\end{equation}
and equality holds when
\begin{equation}
\lambda_{R}=\sqrt{2}\lambda_{L} . \label{equality-strong coupling}
\end{equation}

\begin{figure}
\includegraphics[width=8.6cm]{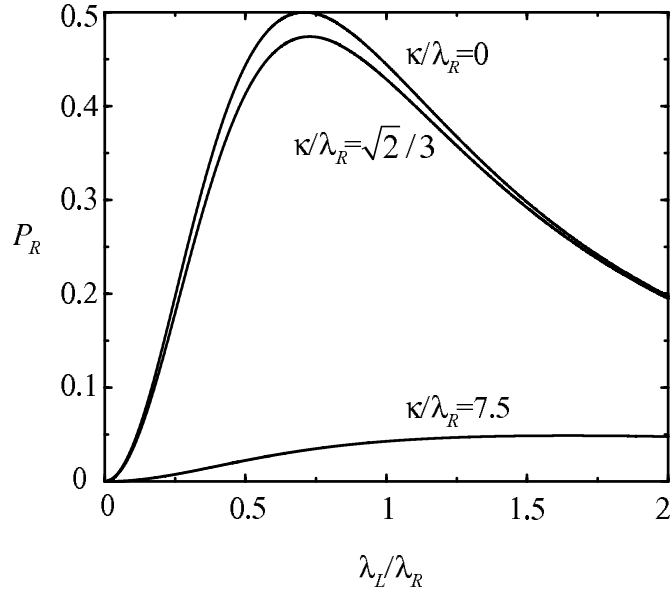}
\caption{A plot of probability of successful generation of the
entangled state, $P_R$, versus $\lambda_L/\lambda_R$ as obtained
from Eq.~(\ref{PR}), with $\delta_e=0$ and $\kappa/\lambda_R=0$,
$\sqrt{2}/3$, and $7.5$.} \label{fig3}
\end{figure}

Figure~\ref{fig3} shows the results of $P_{R}$ versus
$\lambda_{L}/\lambda_{R}$ when $\delta_e =0$, and for different
cavity leakage rates $\kappa/\lambda_R=0$, $\sqrt{2}/3$, and
$7.5$. One can see that the probability depends strongly on the
cavity leakage rate. For finite $\kappa$, $P_{R}$ does not attain
maximum at $\lambda_{R}=\sqrt{2}\lambda_{L}$ exactly and the
maximum value cannot reach $1/2$. In fact, from Eq.~(\ref{P-no
detuning}), we have,
\begin{equation}
P_{R} \leq \frac{2}{\left(
1+\sqrt{1+(\frac{\kappa}{2\lambda_L})^{2}} \right) ^{2}}
<\frac{1}{2}  \label{PRmax}
\end{equation}
and equality holds when
\begin{equation}
\lambda_{R}=\sqrt{2}\left[ 1+\left(
\frac{\kappa}{2\lambda_{L}}\right) ^{2}\right] ^{\frac{1}{4}}
\lambda_L. \label{equality}
\end{equation}

The above results can be better understood by studying the analog
of a ``three-level system" in a perfect cavity. As mentioned
previously, our two-atom model reduces to a single three-level
symmetric system if Eq.~(\ref{equality-strong coupling}) is
satisfied. It is well known that such a symmetric ``three-level
system", initially in one of its ground states, oscillates between
the two ground states. When $\delta_e=0$, the oscillations are
sinusoidal and complete. When there is detuning, the evolution
becomes more complicated and perfect population transfer from one
ground state to the other cannot take place. However, it can be
shown that when the couplings of the two channels are equal
(Eq.~(\ref{equality-strong coupling}) in our model), the
time-averaged populations of the two ground states are equal,
independent of the magnitude of detuning. Moreover, as the cavity
under consideration is leaky, the probability of detecting an $L$
(or $R$) photon outside the cavity is expected to be proportional
to the time the system dwelling in each of the ground states. For
the case of a good cavity, the upper bound $50\%$ of the
probability of success is easy to understand, as the system spends
equal time on both of the ground states in this limit. Hence when
the photon leaks outside the cavity, the atom has equal
probability to decay into an entangled state or a direct product
state.

In Ref.~\cite{Hong}, the hyperfine levels of a cesium atom studied
in Ref.~\cite{Kimble-cesium} were quoted as examples for realistic
implementation of their scheme. Two different sets of atomic
levels are studied. In both cases, $\kappa/\lambda_R=\sqrt{2}/3$,
while $\lambda_L/\lambda_R=1$ or $\sqrt{5/6}$. The numerical
results for $P_R$ they obtained are $0.43$ and $0.45$,
respectively, which agree with our analytical solution shown by
the corresponding curve in Fig.~\ref{fig3}. In fact, from
Eq.~(\ref{P-no detuning}), our analytic results are $P_R=3/7$ and
$9/20$, respectively.

One of the parameters in Fig.~\ref{fig3}, $\kappa/\lambda_R=7.5$,
is taken from a real CQED setup in Ref.~\cite{Kimble-logic-gate}.
The parameters of the setup, when denoted by the corresponding
notations in our model, are given by
$(\lambda_{L(R)},\kappa,\gamma)/2\pi=(20,150,5){\rm MHz}$, where
$\gamma$ is the transverse decay rate of the excited state
\cite{prob}. (The effect of spontaneous decay has so far been
assumed negligible in our study. We shall return to discuss this
point in Sec.~\ref{sub:Input-photon}.) From Fig.~\ref{fig3}, it is
observed that in this case, the Rabi oscillation frequency is not
high enough compared to the cavity leakage rate, the bias due to
the initial state cannot be erased and this lowers the probability
of success for poorer cavities.
As can be concluded from Eq.~(\ref{PRmax}), that in the weak
coupling limit $\lambda_{L(R)} \ll\kappa$, $P_R \simeq 0$, because
there is effectively no oscillation at all before the photon
decays. Hence the scheme using cavity photon is implausible for
weak coupling cases. However, as will be shown in
Sec.~\ref{sub:Input-photon}, the entanglement scheme still works
nicely in the weak coupling regime with an injected photon.

\subsection{Injected photon packet} \label{sub:Input-photon}
The motivation to switch from cavity photons to injected photons
is two-fold. First, it is practically more realistic to consider a
photon injected into the cavity than one that already exists
inside. Second, inspired by the ``three-level system" analogy
discussed in Sec.~\ref{sub:Cavity-photon}, it is believed that an
injected photon with a duration longer than the cavity leakage
time may be able to ``drive" the system into the entangled state
with a higher efficiency. We shall see that by including general
initial photon states other than cavity photon, one may also
eliminate the requirement for strong coupling.

Consider an injected photon with a spectral function
$f(k^{\prime})$ satisfying Eq.~(\ref{scatter-cond}). From
Eqs.~(\ref{scatter-1})-(\ref{scatter-2}), and following similar
arguments outlined in Sec.~\ref{sub:Cavity-photon}, we have
\begin{equation}
\label{psiout-input} |\Psi \rangle _{\rm out} =|LL\rangle \otimes
\int_{-\infty}^{\infty} dk \, e^{-ikt}f(k) C_L(k)
|k_{L}\rangle-|\Phi \rangle \otimes \int_{-\infty}^{\infty} dk \,
e^{-ikt}f(k)C_R(k)|k_{R}\rangle ,
\end{equation}
where
\begin{eqnarray}
C_{L}(k) &=&\frac{(\Delta k-\delta_e)(\Delta
k^{2}+\kappa^{2}/4)-\Delta
k(\lambda_{R}^{2}+2\lambda_{L}^{2})+i\kappa(\lambda_{R}^{2}-2\lambda_{L}^{2})/2}{(\Delta
k-i\kappa/2)(\Delta k-\omega_{+})(\Delta k-\omega_{-})} ,  \label{CL} \\
C_{R}(k) &=&\frac{\sqrt{2}ie^{i\Delta
_{LR}}\kappa\lambda_{L}\lambda_{R}}{(\Delta k-i\kappa/2)(\Delta
k-\omega_{+})(\Delta k-\omega_{-})} . \label{CR}
\end{eqnarray}
It is important to note that perfect transfer occurs
at the photon frequencies
\begin{equation}
\label{Dk-poles} \Delta k=0 ,\, \pm\sqrt{4\lambda_L^2-\kappa^2/4}
\end{equation}
because the corresponding amplitudes,
\begin{eqnarray}
\label{CL=0}
C_L(k)&=&0, \\
\label{CR=1}
C_R(k)&=&e^{i\Delta_{LR}},
\end{eqnarray}
as long as $\delta_e=0$ and Eq.~(\ref{equality-strong coupling})
are satisfied. Therefore, we have an almost unit probability of
generating the entangled state $|\Phi \rangle$ when $f(k)$ is
strongly peaked at one of the frequencies defined in
Eq.~(\ref{Dk-poles}). The efficiency of entanglement generation
depends on the spectral width of the injected photon packet,
compared with the width of the function $C_R(k)$ around the peaks.
Also notice that detuning only shifts the frequencies in
Eq.~(\ref{Dk-poles}), and our scheme remains plausible by
adjusting the peak frequency of the injected photon.

We first study an injected photon packet with a complex Lorentzian
spectral function centered at $k_c$, and with its pole on the
upper half plane:
\begin{equation}
\label{displaced-cavity} f(k^{\prime
})=\lim_{\tau\rightarrow\infty}\frac{\sqrt{\kappa _{\rm in}/2\pi
}}{k^{\prime }-k_{c}-i\kappa _{\rm in}/2}e^{ik^{\prime}\tau}.
\end{equation}
Here the parameter $\kappa_{\rm in}$ is the spectral width of the
input photon packet and $\tau$ determines the initial distance of
the photon packet from the cavity. It can be shown that the
scattering state condition, Eq.~(\ref{scatter-cond}), is satisfied
under the limit $\tau\rightarrow\infty$ in
Eq.~(\ref{displaced-cavity}), and hence the probability of
obtaining the entangled state is given by
\begin{equation}
\label{PR-int} P_R=\int_{-\infty}^{\infty}|f(k)C_R(k)|^2 dk.
\end{equation}
Notice that $P_R$ depends only on the module of the spectral
function $|f(k)|$, hence an injected photon with spectral function
\begin{equation}
\label{input-lorentzian} f(k^{\prime })=\frac{\sqrt{\kappa _{\rm
in}/2\pi}}{k^{\prime }-k_{c}+i\kappa _{\rm in}/2}
\end{equation}
satisfying Eq.~(\ref{scatter-cond}) yields the same probability.
An injected photon with spectral function in the form shown in
Eq.~(\ref{input-lorentzian}) can be obtained, for example, by
injecting the leaked photon emitted by an atom inside another
Fabry-Perot cavity. The number of peaks of the spectral function
depends on the parameters of the cavity and the atom.

\begin{figure}
\includegraphics[width=8.6cm]{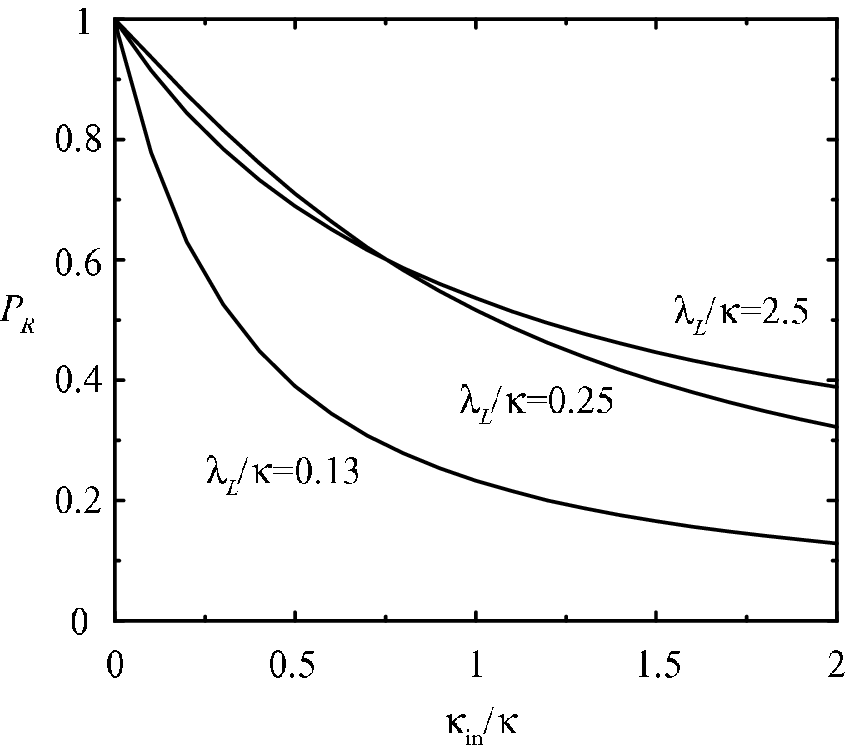}
\caption{A plot of probability of successful generation of the
entangled state, $P_R$, versus $\kappa_{\rm in}/\kappa$ for an
injected photon with a Lorentzian spectral function given by
Eq.~(\ref{input-lorentzian}), with $\delta_e=0$,
$\lambda_L/\kappa=2.5$, $0.25$, and $0.13$.} \label{fig4}
\end{figure}

Figure~\ref{fig4} shows the dependence of $P_R$ on the spectral
width of the injected photon. It is shown that in general our
method can increase the probability of success to well over $50\%$
when the width of the injected photon is sufficiently small
compared with the relaxation rates of the system. The probability
can even approach one in the limiting case of a
quasi-monochromatic injected photon.

We point out that in the strong coupling regime, $\lambda_{L(R)}
\gg \kappa$, the relaxation time of the system is $1/\kappa$.
Therefore a high probability of success can be achieved when
$\kappa_{\rm in}<\kappa$. An example is shown by the
$\lambda_L/\kappa=2.5$ curve in Fig.~\ref{fig4}. On the other
hand, in the bad cavity limit, $\lambda_{L(R)} \ll \kappa$, the
relaxation rate of the system is modified to $(2\lambda_L^2 +
\lambda_R^2)/\kappa$. This explains why the
$\lambda_L/\kappa=0.13$ curve in Fig.~\ref{fig4} requires a
smaller $\kappa_{\rm in}$ in order to obtain a higher probability
of success.

It is useful to note that the curve in Fig.~\ref{fig4} with
$\lambda_L/\kappa=0.13$ corresponds to a realistic situation using
the parameters in the setup in Ref.~\cite{Kimble-logic-gate}.  For
the cavity photon scheme discussed in the previous section, the
probability of success is less than $5\%$. However, using our
injected photon scheme, a probability near $50\%$ can still be
obtained even if $\kappa_{\rm in}/\kappa \sim 0.3$ (see
Fig.~\ref{fig4}).

From Fig.~\ref{fig4}, one observes that an intermediate strength
of coupling $\lambda_L/\kappa=0.25$ yields best results. This
particular value of the parameter is chosen so that the three
roots in Eq.~(\ref{Dk-poles}) coincide. From Eq.~(\ref{CL}), one
sees that if the three roots are far separated, the behavior of
$C_L(k)$ around $k_c$ is linear: $C_L(k)\sim O(\Delta k)$.
However, when the three roots coincide, the behavior becomes
cubic: $C_L(k)\sim O(\Delta k)^3$. As the requirement of energy
conservation leads directly to $|C_L(k)|^2+|C_R(k)|^2=1$, this
means $|C_R(k)|$ is close to one for a wide range of frequencies
near $\Delta k=0$. For the specific form of the injected photon we
assumed, which peaks at the resonance frequency, this means even a
larger spectral width is allowed to yield a satisfactory
efficiency.

The significance of the spectral shape of the injected photon
becomes more apparent for a Gaussian packet. Consider
\begin{equation}
\label{input-gaussian} f(k^{\prime
})=\lim_{\tau\rightarrow\infty}\frac{\sqrt{2}}{\pi^{1/4}\sqrt{\kappa_{\rm
in}}}\, \exp\left[-\frac{2(k^{\prime}-k_c)^2}{\kappa_{\rm
in}^2}+ik^{\prime}\tau\right],
\end{equation}
which represents a photon packet with peak frequency $k_c$ and
spectral width $\kappa_{\rm in}$. Notice that
Eq.~(\ref{scatter-cond}) is satisfied owing to the phase factor
$e^{ik^{\prime}\tau}$, and under the limit
$\tau\rightarrow\infty$. Hence the probability of obtaining the
entangled state is again given by Eq.~(\ref{PR-int}).

\begin{figure}
\includegraphics[width=8.6cm]{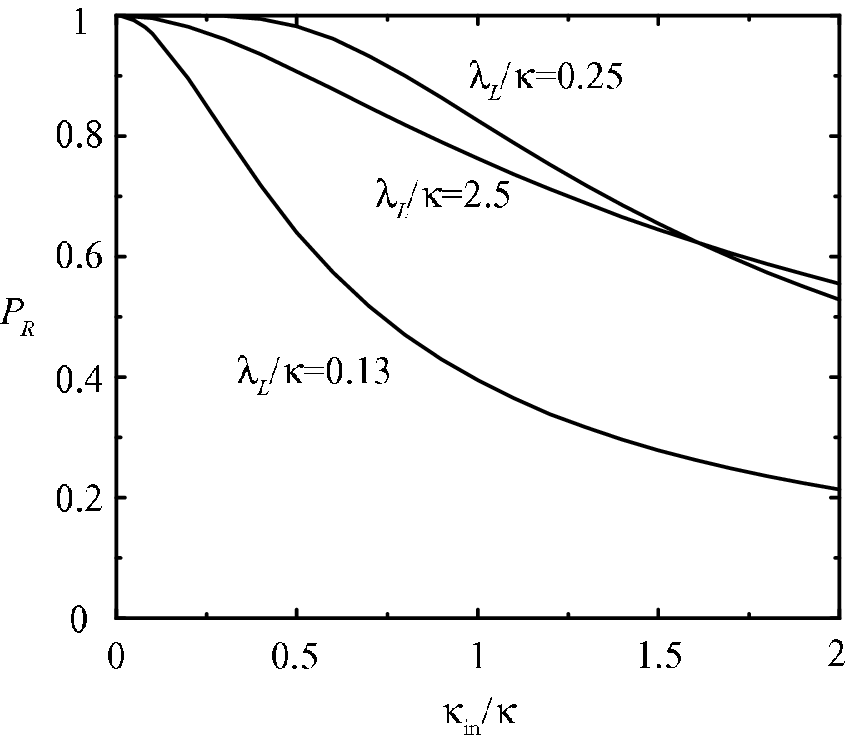}
\caption{A plot of probability of successful generation of the
entangled state, $P_R$, versus $\kappa_{\rm in}/\kappa$ for an
injected photon with a Gaussian spectral function given by
Eq.~(\ref{input-gaussian}), with $\delta_e=0$,
$\lambda_L/\kappa=2.5$, $0.25$ and $0.13$.} \label{fig5}
\end{figure}

Figure~\ref{fig5} shows the dependence of $P_R$ on the width of
the injected photon with the same set of parameters as in the
study of Lorentzian spectral function. One can conclude that in
general the efficiency is improved using a packet with a Gaussian
spectral function. This can be understood because for the same
spectral width, a Gaussian spectral function is more concentrated
around the peak frequency than that of a Lorentzian. Remarkably,
notice that in the case $\lambda_L/\kappa=0.25$, $P_R\sim 1$ even
for relatively large spectral widths. In fact, we can have
$P_R>0.999$ for $\kappa_{\rm in}\sim 0.3\kappa$ in a single
operation, which can essentially be considered deterministic.

Finally, we would like to address the effects of spontaneous
decay. The coupling of the atoms to non-cavity modes broadens the
excited-levels. This effect can be readily included by introducing
an imaginary part to the energy of the atomic excited states. In
other words, we only have to replace $\omega_e$ by
$\omega_e-i\gamma/2$, and all the equations remain valid. For
cesium systems such as that in Ref.~\cite{Kimble-cesium}, we found
that the presence of a non-zero $\gamma$ only causes a minute
decrease of the probability. For example, by employing the
hyperfine levels $|L\rangle=|6S_{1/2},F=4,m_F=4\rangle$,
$|R\rangle=|6S_{1/2},F=4,m_F=2\rangle$, and
$|e\rangle=|6P_{1/2},F=4,m_F=3\rangle$, we have
$\lambda_R/\lambda_L=\sqrt{7/4}$. With $\gamma=0.033\kappa$ and
$\lambda_L=0.25\kappa$, one can still obtain a value of $P_R$ near
$93\%$ using an injected Gaussian photon packet with $\kappa_{\rm
in}\sim 0.3\kappa$. Therefore, the current scheme is quite robust
against spontaneous atomic decay.

\section{Conclusion} \label{sec:Conclusion}
In conclusion, we study in detail a scheme proposed recently to
generate entangled states of two identical $\Lambda$-type
three-level atoms inside a leaky cavity. The atoms are initially
prepared in the ground states on the same side of the $\Lambda$
systems, with the presence of a single photon, which is either in
a cavity mode or injected from the exterior of the cavity. First,
for a cavity photon, we show analytically that an entangled state
can be generated in a single trial with a certain probability
always less than $50\%$, corroborating the numerical result
obtained in Ref.~\cite{Hong}. However, their scheme is plausible
only in the strong coupling regime. By drawing an analogy with a
``three-level system", we provide an intuitive understanding of
these results. Second, for an injected photon, we show that the
probability can be increased by injecting a spectrally narrow
photon. In particular, we show that an almost unit probability can
be achieved with a Gaussian packet by exploiting the conditions
mentioned above.

Compared with the scheme proposed in Ref.~\cite{Hong}, which
remains probabilistic for each single trial, and requires a
feedback mechanism to achieve a unit probability, our proposal
here can surely increase the success rate in each trial and hence
significantly reduces the number of trials required to yield an
entangled state. We have therefore presented here a feasible
scheme to generate the maximally entangled state of two
$\Lambda$-type three-level atoms with a high probability, which
can effectively be classified as ``deterministic". This is also in
contrast to recent probabilistic schemes, in which intrinsic
uncertainties are inherent in quantum measurement processes
\cite{Probabilistic-entangle-cavity,Duan-multiatom-entangle}.
However, the main challenge of realizing our scheme is the
requirement of a single photon source, which is now under active
investigations and may become feasible in the near future
\cite{Brattke}.

\begin{acknowledgments}
We thank H.~T.~Fung for discussions. The work described in this
paper is partially supported by two grants from the Research
Grants Council of the Hong Kong Special Administrative Region,
China (Project Nos. 428200 and 401603) and a direct grant (Project
No. 2060150) from the Chinese University of Hong Kong.
\end{acknowledgments}

\appendix
\section{}
In our model, we are interested in initial photon states that may
or may not be obtained by using Eqs.~(\ref{scatter-1}). For
example, it can be shown that with a cavity photon defined by
Eq.~(\ref{cavity-mode}), or a Gaussian packet defined by
\begin{equation}
f(k^{\prime })=\frac{\sqrt{2}}{\pi^{1/4}\sqrt{\kappa_{\rm in}}}\,
\exp\left[-\frac{2(k^{\prime}-k_c)^2}{\kappa_{\rm in}^2}\right],
\end{equation}
Eqs.~(\ref{Gkk'}) and (\ref{scatter-1}) lead to different results.
However, by introducing a phase factor in the spectral functions,
Eq.~(\ref{scatter-1}) becomes valid for a Gaussian packet defined
by Eq.~(\ref{input-gaussian}) or a ``displaced" cavity photon in
Eq.~(\ref{displaced-cavity}). In addition, for some packets
without a phase factor like $e^{ik^{\prime}\tau}$, such as a
complex Lorentzian defined by Eq.~(\ref{input-lorentzian}),
Eq.~(\ref{scatter-1}) still yields the correct results. Hence, it
is desirable to obtain a necessary and sufficient condition for
the applicability of Eq.~(\ref{scatter-1}). Here we will prove
that Eq.~(\ref{scatter-cond}) is the condition we are looking for.

For the initial state
\begin{equation}
|\Psi\rangle_{\rm in}=\int_{-\infty}^{\infty}
dk^{\prime}f(k^{\prime})|\psi_1(k^{\prime})\rangle,
\end{equation}
Eq.~(\ref{scatter-cond}) leads to
\begin{equation}
\label{VPsi=0}
\begin{array}{ccccc}
\int_{-\infty}^{\infty}
f(k^{\prime})V(k^{\prime})e^{-ik^{\prime}t}dk^{\prime}=0 & & &
{\rm for} & t \le 0.
\end{array}
\end{equation}
Equation~(\ref{VPsi=0}) implies
\begin{equation}
\int_{-\infty}^{\infty} dt \int_{-\infty}^{\infty}
dk^{\prime}f(k^{\prime})V(k^{\prime})e^{i(z-k^{\prime})t}=\int_{0}^{\infty}
dt \int_{-\infty}^{\infty}
dk^{\prime}f(k^{\prime})V(k^{\prime})e^{i(z-k^{\prime})t}.
\end{equation}
The proper meaning of the above equation must be defined by
certain limiting process. For our problem, we choose to introduce
to the real number $z$ a small imaginary part, which has the same
sign as $t$. Hence
\begin{equation}
\lim_{\epsilon\rightarrow0^+}\int_{0}^{\infty} dt
\int_{-\infty}^{\infty}
dk^{\prime}f(k^{\prime})V(k^{\prime})e^{(iz-ik^{\prime}-\epsilon)t}=
\lim_{\epsilon\rightarrow0^+}\int_{-\infty}^{\infty} dt
\int_{-\infty}^{\infty}
dk^{\prime}f(k^{\prime})V(k^{\prime})e^{(iz-ik^{\prime}-{\rm
sgn}(t)\epsilon)t}.
\end{equation}
Therefore
\begin{equation}
\lim_{\epsilon\rightarrow0^+}\int_{-\infty}^{\infty}
f(k^{\prime})V(k^{\prime})\frac{i}{z+i\epsilon-k^{\prime}}dk^{\prime}=2\pi\int_{-\infty}^{\infty}
f(k^{\prime})V(k^{\prime})\delta(z-k^{\prime})dk^{\prime},
\end{equation}
or
\begin{equation}
\label{scatter-cond-2}
\lim_{\epsilon\rightarrow0^+}\int_{-\infty}^{\infty}
\frac{f(k^{\prime})V(k^{\prime})}{z+i\epsilon-k^{\prime}}dk^{\prime}
=-2\pi i f(z)V(z).
\end{equation}
From Eq.~(\ref{Gkk'}), we can conclude that
\begin{eqnarray}
\langle \psi_1(k)|\hat{G}(\omega)|\Psi\rangle_{\rm
in}&=&\int_{-\infty}^{\infty}\left\{\frac{\delta(k-k^{\prime})}{\omega-k}
+\frac{V(k^{\prime})V^\ast(k)}{(\omega-k)(\omega-k^{\prime})}\langle
E|\hat{G}(\omega)|E\rangle\right\}f(k^{\prime}) dk^{\prime}\\
&=&\frac{f(k)}{\omega-k}+\frac{V(k)}{\omega-k}\langle
E|\hat{G}(\omega)|E\rangle\int_{-\infty}^{\infty}
\frac{f(k^{\prime})V(k^{\prime})}{\omega-k^{\prime}}dk^{\prime}
\end{eqnarray}
For $t\ge 0$, the evolution is governed by the retarded
propagator. We can thus let $\omega$ approach the real axis from
above, and from Eq.~(\ref{scatter-cond-2}), we have
\begin{equation}
\langle \psi_1(k)|\hat{G}(\omega)|\Psi\rangle_{\rm
in}=\frac{f(k)}{\omega-k}-2\pi i
\frac{f(\omega)V(\omega)V(k)\langle
E|\hat{G}(\omega)|E\rangle}{\omega-k}.
\end{equation}
Because $f(k)$ represents a wave-packet, $\hat{V}$ has finite
support in space, and $\langle E|\hat{G}(\omega)|E\rangle$ governs
the decaying process of the excited state, $k$ is the only real
pole that one needs to keep in the asymptotic long time limit.
Hence by performing the inverse Fourier transform, we have
\begin{equation}
\langle \psi_1(k)|\hat{U}(t,0)|\Psi\rangle_{\rm
in}=f(k)e^{-ikt}\left(1-2\pi i V(k)^2\langle
E|\hat{G}(k)|E\rangle\right),
\end{equation}
which is equivalent to defining the $S$-matrix as shown in
Eq.~(\ref{scatter-1}).

\bibliography{myref}

\begin{thebibliography}{40}
\expandafter\ifx\csname natexlab\endcsname\relax\def\natexlab#1{#1}\fi
\expandafter\ifx\csname bibnamefont\endcsname\relax
  \def\bibnamefont#1{#1}\fi
\expandafter\ifx\csname bibfnamefont\endcsname\relax
  \def\bibfnamefont#1{#1}\fi
\expandafter\ifx\csname citenamefont\endcsname\relax
  \def\citenamefont#1{#1}\fi
\expandafter\ifx\csname url\endcsname\relax
  \def\url#1{\texttt{#1}}\fi
\expandafter\ifx\csname urlprefix\endcsname\relax\def\urlprefix{URL }\fi
\providecommand{\bibinfo}[2]{#2}
\providecommand{\eprint}[2][]{\url{#2}}

\bibitem[{\citenamefont{Schr{\"o}dinger}(1935)}]{Schrodinger-Entanglement}
\bibinfo{author}{\bibfnamefont{E.}~\bibnamefont{Schr{\"o}dinger}},
  \bibinfo{journal}{Naturwissenschaften} \textbf{\bibinfo{volume}{23}},
  \bibinfo{pages}{807} (\bibinfo{year}{1935}).

\bibitem[{\citenamefont{Einstein et~al.}(1935)\citenamefont{Einstein, Podolsky,
  and Rosen}}]{EPR}
\bibinfo{author}{\bibfnamefont{A.}~\bibnamefont{Einstein}},
  \bibinfo{author}{\bibfnamefont{B.}~\bibnamefont{Podolsky}}, \bibnamefont{and}
  \bibinfo{author}{\bibfnamefont{N.}~\bibnamefont{Rosen}},
  \bibinfo{journal}{Phys. Rev.} \textbf{\bibinfo{volume}{47}},
  \bibinfo{pages}{777} (\bibinfo{year}{1935}).

\bibitem[{\citenamefont{Bell}(1965)}]{Bell}
\bibinfo{author}{\bibfnamefont{J.~S.} \bibnamefont{Bell}},
  \bibinfo{journal}{Physics (Long Island City, NY)}
  \textbf{\bibinfo{volume}{1}}, \bibinfo{pages}{195} (\bibinfo{year}{1965}).

\bibitem[{\citenamefont{Bouwmeester et~al.}(2000)\citenamefont{Bouwmeester,
  Ekert, and Zeilinger}}]{Bo-book}
\bibinfo{editor}{\bibfnamefont{D.}~\bibnamefont{Bouwmeester}},
  \bibinfo{editor}{\bibfnamefont{A.}~\bibnamefont{Ekert}}, \bibnamefont{and}
  \bibinfo{editor}{\bibfnamefont{A.}~\bibnamefont{Zeilinger}}, eds.,
  \emph{\bibinfo{title}{The Physics of Quantum Information}}
  (\bibinfo{publisher}{Springer-Verlag}, \bibinfo{address}{Berlin},
  \bibinfo{year}{2000}).

\bibitem[{\citenamefont{Nielsen and Chuang}(2000)}]{Chuang-book}
\bibinfo{author}{\bibfnamefont{M.~A.} \bibnamefont{Nielsen}} \bibnamefont{and}
  \bibinfo{author}{\bibfnamefont{I.~L.} \bibnamefont{Chuang}},
  \emph{\bibinfo{title}{Quantum Computation and Quantum Information}}
  (\bibinfo{publisher}{Cambridge University Press},
  \bibinfo{address}{Cambridge, England}, \bibinfo{year}{2000}).

\bibitem[{\citenamefont{Raimond et~al.}(2001)\citenamefont{Raimond, Brune, and
  Haroche}}]{Haroche-entanglement-RMP}
\bibinfo{author}{\bibfnamefont{J.~M.} \bibnamefont{Raimond}},
  \bibinfo{author}{\bibfnamefont{M.}~\bibnamefont{Brune}}, \bibnamefont{and}
  \bibinfo{author}{\bibfnamefont{S.}~\bibnamefont{Haroche}},
  \bibinfo{journal}{Rev. Mod. Phys.} \textbf{\bibinfo{volume}{73}},
  \bibinfo{pages}{565} (\bibinfo{year}{2001}).

\bibitem[{\citenamefont{Bouchiat et~al.}(1999)\citenamefont{Bouchiat, Vion,
  Joyez, Est\`eve, and Devoret}}]{Bouchiat}
\bibinfo{author}{\bibfnamefont{V.}~\bibnamefont{Bouchiat}},
  \bibinfo{author}{\bibfnamefont{D.}~\bibnamefont{Vion}},
  \bibinfo{author}{\bibfnamefont{P.}~\bibnamefont{Joyez}},
  \bibinfo{author}{\bibfnamefont{D.}~\bibnamefont{Est\`eve}}, \bibnamefont{and}
  \bibinfo{author}{\bibfnamefont{M.~H.} \bibnamefont{Devoret}},
  \bibinfo{journal}{J. Supercond.} \textbf{\bibinfo{volume}{12}},
  \bibinfo{pages}{789} (\bibinfo{year}{1999}).

\bibitem[{\citenamefont{Bertoni et~al.}(2000)\citenamefont{Bertoni, Bordone,
  Brunetti, Jacoboni, and Reggiani}}]{Bertoni}
\bibinfo{author}{\bibfnamefont{A.}~\bibnamefont{Bertoni}},
  \bibinfo{author}{\bibfnamefont{P.}~\bibnamefont{Bordone}},
  \bibinfo{author}{\bibfnamefont{R.}~\bibnamefont{Brunetti}},
  \bibinfo{author}{\bibfnamefont{C.}~\bibnamefont{Jacoboni}}, \bibnamefont{and}
  \bibinfo{author}{\bibfnamefont{S.}~\bibnamefont{Reggiani}},
  \bibinfo{journal}{Phys. Rev. Lett.} \textbf{\bibinfo{volume}{84}},
  \bibinfo{pages}{5912} (\bibinfo{year}{2000}).

\bibitem[{\citenamefont{Friedman et~al.}(2000)\citenamefont{Friedman, Patel,
  Chen, Tolpygo, and Lukens}}]{Friedman}
\bibinfo{author}{\bibfnamefont{J.~R.} \bibnamefont{Friedman}},
  \bibinfo{author}{\bibfnamefont{V.}~\bibnamefont{Patel}},
  \bibinfo{author}{\bibfnamefont{W.}~\bibnamefont{Chen}},
  \bibinfo{author}{\bibfnamefont{S.~K.} \bibnamefont{Tolpygo}},
  \bibnamefont{and} \bibinfo{author}{\bibfnamefont{J.~E.}
  \bibnamefont{Lukens}}, \bibinfo{journal}{Nature (London)}
  \textbf{\bibinfo{volume}{406}}, \bibinfo{pages}{43} (\bibinfo{year}{2000}).

\bibitem[{\citenamefont{van~der Wal et~al.}(2000)\citenamefont{van~der Wal, ter
  Haar, Wilhelm, Schouten, Harmans, Orlando, Lloyd, and Mooij}}]{Lloyd-sci-2}
\bibinfo{author}{\bibfnamefont{C.~H.} \bibnamefont{van~der Wal}},
  \bibinfo{author}{\bibfnamefont{A.~C.~J.} \bibnamefont{ter Haar}},
  \bibinfo{author}{\bibfnamefont{F.~K.} \bibnamefont{Wilhelm}},
  \bibinfo{author}{\bibfnamefont{R.~N.} \bibnamefont{Schouten}},
  \bibinfo{author}{\bibfnamefont{C.~J. P.~M.} \bibnamefont{Harmans}},
  \bibinfo{author}{\bibfnamefont{T.~P.} \bibnamefont{Orlando}},
  \bibinfo{author}{\bibfnamefont{S.}~\bibnamefont{Lloyd}}, \bibnamefont{and}
  \bibinfo{author}{\bibfnamefont{J.~E.} \bibnamefont{Mooij}},
  \bibinfo{journal}{Science} \textbf{\bibinfo{volume}{290}},
  \bibinfo{pages}{773} (\bibinfo{year}{2000}).

\bibitem[{\citenamefont{Gershenfeld and Chuang}(1997)}]{Chuang-sci}
\bibinfo{author}{\bibfnamefont{N.~A.} \bibnamefont{Gershenfeld}}
  \bibnamefont{and} \bibinfo{author}{\bibfnamefont{I.~L.}
  \bibnamefont{Chuang}}, \bibinfo{journal}{Science}
  \textbf{\bibinfo{volume}{275}}, \bibinfo{pages}{350} (\bibinfo{year}{1997}).

\bibitem[{\citenamefont{Jones et~al.}(1998)\citenamefont{Jones, Mosca, and
  Hansen}}]{Jones-nature}
\bibinfo{author}{\bibfnamefont{J.~A.} \bibnamefont{Jones}},
  \bibinfo{author}{\bibfnamefont{M.}~\bibnamefont{Mosca}}, \bibnamefont{and}
  \bibinfo{author}{\bibfnamefont{R.~H.} \bibnamefont{Hansen}},
  \bibinfo{journal}{Nature (London)} \textbf{\bibinfo{volume}{393}},
  \bibinfo{pages}{344} (\bibinfo{year}{1998}).

\bibitem[{\citenamefont{Monroe et~al.}(1995)\citenamefont{Monroe, Meekhof,
  King, Itano, and Wineland}}]{Monroe-gate}
\bibinfo{author}{\bibfnamefont{C.}~\bibnamefont{Monroe}},
  \bibinfo{author}{\bibfnamefont{D.~M.} \bibnamefont{Meekhof}},
  \bibinfo{author}{\bibfnamefont{B.~E.} \bibnamefont{King}},
  \bibinfo{author}{\bibfnamefont{W.~M.} \bibnamefont{Itano}}, \bibnamefont{and}
  \bibinfo{author}{\bibfnamefont{D.~J.} \bibnamefont{Wineland}},
  \bibinfo{journal}{Phys. Rev. Lett.} \textbf{\bibinfo{volume}{75}},
  \bibinfo{pages}{4714} (\bibinfo{year}{1995}).

\bibitem[{\citenamefont{Turchette et~al.}(1998)\citenamefont{Turchette, Wood,
  King, Myatt, Leibfried, Itano, Monroe, and
  Wineland}}]{Turchette-entangle-trap}
\bibinfo{author}{\bibfnamefont{Q.~A.} \bibnamefont{Turchette}},
  \bibinfo{author}{\bibfnamefont{C.~S.} \bibnamefont{Wood}},
  \bibinfo{author}{\bibfnamefont{B.~E.} \bibnamefont{King}},
  \bibinfo{author}{\bibfnamefont{C.~J.} \bibnamefont{Myatt}},
  \bibinfo{author}{\bibfnamefont{D.}~\bibnamefont{Leibfried}},
  \bibinfo{author}{\bibfnamefont{W.~M.} \bibnamefont{Itano}},
  \bibinfo{author}{\bibfnamefont{C.}~\bibnamefont{Monroe}}, \bibnamefont{and}
  \bibinfo{author}{\bibfnamefont{D.~J.} \bibnamefont{Wineland}},
  \bibinfo{journal}{Phys. Rev. Lett.} \textbf{\bibinfo{volume}{81}},
  \bibinfo{pages}{3631} (\bibinfo{year}{1998}).

\bibitem[{\citenamefont{Sackett et~al.}(2000)\citenamefont{Sackett, Kielpinski,
  King, Langer, Meyer, Myatt, Rowe, Turchette, Itano, Wienland
  et~al.}}]{Sackett-nature}
\bibinfo{author}{\bibfnamefont{C.~A.} \bibnamefont{Sackett}},
  \bibinfo{author}{\bibfnamefont{D.}~\bibnamefont{Kielpinski}},
  \bibinfo{author}{\bibfnamefont{B.~E.} \bibnamefont{King}},
  \bibinfo{author}{\bibfnamefont{C.}~\bibnamefont{Langer}},
  \bibinfo{author}{\bibfnamefont{V.}~\bibnamefont{Meyer}},
  \bibinfo{author}{\bibfnamefont{C.~J.} \bibnamefont{Myatt}},
  \bibinfo{author}{\bibfnamefont{M.}~\bibnamefont{Rowe}},
  \bibinfo{author}{\bibfnamefont{Q.~A.} \bibnamefont{Turchette}},
  \bibinfo{author}{\bibfnamefont{W.~M.} \bibnamefont{Itano}},
  \bibinfo{author}{\bibfnamefont{D.~J.} \bibnamefont{Wienland}},
  \bibnamefont{et~al.}, \bibinfo{journal}{Nature (London)}
  \textbf{\bibinfo{volume}{404}}, \bibinfo{pages}{256} (\bibinfo{year}{2000}).

\bibitem[{\citenamefont{Yamamoto et~al.}(2003)\citenamefont{Yamamoto, Koashi,
  Ozdemir, and Imoto}}]{Yamamoto-nature}
\bibinfo{author}{\bibfnamefont{T.}~\bibnamefont{Yamamoto}},
  \bibinfo{author}{\bibfnamefont{M.}~\bibnamefont{Koashi}},
  \bibinfo{author}{\bibfnamefont{S.~K.} \bibnamefont{Ozdemir}},
  \bibnamefont{and} \bibinfo{author}{\bibfnamefont{N.}~\bibnamefont{Imoto}},
  \bibinfo{journal}{Nature (London)} \textbf{\bibinfo{volume}{421}},
  \bibinfo{pages}{343} (\bibinfo{year}{2003}).

\bibitem[{\citenamefont{Abouraddy et~al.}(2001)\citenamefont{Abouraddy, Saleh,
  Sergienko, and Teich}}]{two-photon-imaging}
\bibinfo{author}{\bibfnamefont{A.~F.} \bibnamefont{Abouraddy}},
  \bibinfo{author}{\bibfnamefont{B.~E.~A.} \bibnamefont{Saleh}},
  \bibinfo{author}{\bibfnamefont{A.~V.} \bibnamefont{Sergienko}},
  \bibnamefont{and} \bibinfo{author}{\bibfnamefont{M.~C.} \bibnamefont{Teich}},
  \bibinfo{journal}{Phys. Rev. Lett.} \textbf{\bibinfo{volume}{87}},
  \bibinfo{pages}{123602} (\bibinfo{year}{2001}).

\bibitem[{\citenamefont{Lange and Kimble}(2000)}]{Kimble-cesium}
\bibinfo{author}{\bibfnamefont{W.}~\bibnamefont{Lange}} \bibnamefont{and}
  \bibinfo{author}{\bibfnamefont{H.~J.} \bibnamefont{Kimble}},
  \bibinfo{journal}{Phys. Rev. A} \textbf{\bibinfo{volume}{61}},
  \bibinfo{pages}{063817} (\bibinfo{year}{2000}).

\bibitem[{\citenamefont{Rauschenbeutel
  et~al.}(1999)\citenamefont{Rauschenbeutel, Nogues, Osnaghi, Bertet, Brune,
  Raimond, and Haroche}}]{Haroche-phase-gate}
\bibinfo{author}{\bibfnamefont{A.}~\bibnamefont{Rauschenbeutel}},
  \bibinfo{author}{\bibfnamefont{G.}~\bibnamefont{Nogues}},
  \bibinfo{author}{\bibfnamefont{S.}~\bibnamefont{Osnaghi}},
  \bibinfo{author}{\bibfnamefont{P.}~\bibnamefont{Bertet}},
  \bibinfo{author}{\bibfnamefont{M.}~\bibnamefont{Brune}},
  \bibinfo{author}{\bibfnamefont{J.~M.} \bibnamefont{Raimond}},
  \bibnamefont{and} \bibinfo{author}{\bibfnamefont{S.}~\bibnamefont{Haroche}},
  \bibinfo{journal}{Phys. Rev. Lett.} \textbf{\bibinfo{volume}{83}},
  \bibinfo{pages}{5166} (\bibinfo{year}{1999}).

\bibitem[{\citenamefont{Ma{\^ i}tre et~al.}(1997)\citenamefont{Ma{\^ i}tre,
  Hagley, Nogues, Wunderlich, Goy, Brune, Raimond, and
  Haroche}}]{Memory-photon}
\bibinfo{author}{\bibfnamefont{X.}~\bibnamefont{Ma{\^ i}tre}},
  \bibinfo{author}{\bibfnamefont{E.}~\bibnamefont{Hagley}},
  \bibinfo{author}{\bibfnamefont{G.}~\bibnamefont{Nogues}},
  \bibinfo{author}{\bibfnamefont{C.}~\bibnamefont{Wunderlich}},
  \bibinfo{author}{\bibfnamefont{P.}~\bibnamefont{Goy}},
  \bibinfo{author}{\bibfnamefont{M.}~\bibnamefont{Brune}},
  \bibinfo{author}{\bibfnamefont{J.~M.} \bibnamefont{Raimond}},
  \bibnamefont{and} \bibinfo{author}{\bibfnamefont{S.}~\bibnamefont{Haroche}},
  \bibinfo{journal}{Phys. Rev. Lett.} \textbf{\bibinfo{volume}{79}},
  \bibinfo{pages}{769} (\bibinfo{year}{1997}).

\bibitem[{\citenamefont{Plenio et~al.}(1999)\citenamefont{Plenio, Huelga,
  Beige, and Knight}}]{Plenio}
\bibinfo{author}{\bibfnamefont{M.~B.} \bibnamefont{Plenio}},
  \bibinfo{author}{\bibfnamefont{S.~F.} \bibnamefont{Huelga}},
  \bibinfo{author}{\bibfnamefont{A.}~\bibnamefont{Beige}}, \bibnamefont{and}
  \bibinfo{author}{\bibfnamefont{P.~L.} \bibnamefont{Knight}},
  \bibinfo{journal}{Phys. Rev. A} \textbf{\bibinfo{volume}{59}},
  \bibinfo{pages}{2468} (\bibinfo{year}{1999}).

\bibitem[{\citenamefont{Beige et~al.}(2000)\citenamefont{Beige, Braun,
  Tregenna, and Knight}}]{Beige}
\bibinfo{author}{\bibfnamefont{A.}~\bibnamefont{Beige}},
  \bibinfo{author}{\bibfnamefont{D.}~\bibnamefont{Braun}},
  \bibinfo{author}{\bibfnamefont{B.}~\bibnamefont{Tregenna}}, \bibnamefont{and}
  \bibinfo{author}{\bibfnamefont{P.~L.} \bibnamefont{Knight}},
  \bibinfo{journal}{Phys. Rev. Lett.} \textbf{\bibinfo{volume}{85}},
  \bibinfo{pages}{1762} (\bibinfo{year}{2000}).

\bibitem[{\citenamefont{Pachos and Walther}(2002)}]{Pachos}
\bibinfo{author}{\bibfnamefont{J.}~\bibnamefont{Pachos}} \bibnamefont{and}
  \bibinfo{author}{\bibfnamefont{H.}~\bibnamefont{Walther}},
  \bibinfo{journal}{Phys. Rev. Lett.} \textbf{\bibinfo{volume}{89}},
  \bibinfo{pages}{187903} (\bibinfo{year}{2002}).

\bibitem[{\citenamefont{S{\o}rensen and
  M{\o}lmer}(2003)}]{Probabilistic-entangle-cavity}
\bibinfo{author}{\bibfnamefont{A.~S.} \bibnamefont{S{\o}rensen}}
  \bibnamefont{and}
  \bibinfo{author}{\bibfnamefont{K.}~\bibnamefont{M{\o}lmer}},
  \bibinfo{journal}{Phys. Rev. Lett.} \textbf{\bibinfo{volume}{90}},
  \bibinfo{pages}{127903} (\bibinfo{year}{2003}).

\bibitem[{\citenamefont{Hagley et~al.}(1997)\citenamefont{Hagley, Maitre,
  Nogues, Wunderlich, Brune, Raimond, and Haroche}}]{Haroche-EPR}
\bibinfo{author}{\bibfnamefont{E.}~\bibnamefont{Hagley}},
  \bibinfo{author}{\bibfnamefont{X.}~\bibnamefont{Maitre}},
  \bibinfo{author}{\bibfnamefont{G.}~\bibnamefont{Nogues}},
  \bibinfo{author}{\bibfnamefont{C.}~\bibnamefont{Wunderlich}},
  \bibinfo{author}{\bibfnamefont{M.}~\bibnamefont{Brune}},
  \bibinfo{author}{\bibfnamefont{J.~M.} \bibnamefont{Raimond}},
  \bibnamefont{and} \bibinfo{author}{\bibfnamefont{S.}~\bibnamefont{Haroche}},
  \bibinfo{journal}{Phys. Rev. Lett.} \textbf{\bibinfo{volume}{79}},
  \bibinfo{pages}{1} (\bibinfo{year}{1997}).

\bibitem[{\citenamefont{Duan and Kimble}(2003)}]{Duan-multiatom-entangle}
\bibinfo{author}{\bibfnamefont{L.-M.} \bibnamefont{Duan}} \bibnamefont{and}
  \bibinfo{author}{\bibfnamefont{H.~J.} \bibnamefont{Kimble}},
  \bibinfo{journal}{Phys. Rev. Lett.} \textbf{\bibinfo{volume}{90}},
  \bibinfo{pages}{253601} (\bibinfo{year}{2003}).

\bibitem[{\citenamefont{Braun}(2002)}]{Braun}
\bibinfo{author}{\bibfnamefont{D.}~\bibnamefont{Braun}},
  \bibinfo{journal}{Phys. Rev. Lett.} \textbf{\bibinfo{volume}{89}},
  \bibinfo{pages}{277901} (\bibinfo{year}{2002}).

\bibitem[{\citenamefont{Hong and Lee}(2002)}]{Hong}
\bibinfo{author}{\bibfnamefont{J.}~\bibnamefont{Hong}} \bibnamefont{and}
  \bibinfo{author}{\bibfnamefont{H.~W.} \bibnamefont{Lee}},
  \bibinfo{journal}{Phys. Rev. Lett.} \textbf{\bibinfo{volume}{89}},
  \bibinfo{pages}{237901} (\bibinfo{year}{2002}).

\bibitem[{\citenamefont{Ley and Loudon}(1987)}]{Loudon}
\bibinfo{author}{\bibfnamefont{M.}~\bibnamefont{Ley}} \bibnamefont{and}
  \bibinfo{author}{\bibfnamefont{R.}~\bibnamefont{Loudon}},
  \bibinfo{journal}{J. Mod. Opt.} \textbf{\bibinfo{volume}{34}},
  \bibinfo{pages}{227} (\bibinfo{year}{1987}).

\bibitem[{\citenamefont{Lai et~al.}(1988)\citenamefont{Lai, Leung, and
  Young}}]{Lai-narrow-resonance}
\bibinfo{author}{\bibfnamefont{H.~M.} \bibnamefont{Lai}},
  \bibinfo{author}{\bibfnamefont{P.~T.} \bibnamefont{Leung}}, \bibnamefont{and}
  \bibinfo{author}{\bibfnamefont{K.}~\bibnamefont{Young}},
  \bibinfo{journal}{Phys. Rev. A} \textbf{\bibinfo{volume}{37}},
  \bibinfo{pages}{1597} (\bibinfo{year}{1988}).

\bibitem[{\citenamefont{Gea-Banacloche
  et~al.}(1990)\citenamefont{Gea-Banacloche, Lu, Pedrotti, Prasad, Scully, and
  W\'odkiewicz}}]{Scully-MOU}
\bibinfo{author}{\bibfnamefont{J.}~\bibnamefont{Gea-Banacloche}},
  \bibinfo{author}{\bibfnamefont{N.}~\bibnamefont{Lu}},
  \bibinfo{author}{\bibfnamefont{L.~M.} \bibnamefont{Pedrotti}},
  \bibinfo{author}{\bibfnamefont{S.}~\bibnamefont{Prasad}},
  \bibinfo{author}{\bibfnamefont{M.~O.} \bibnamefont{Scully}},
  \bibnamefont{and}
  \bibinfo{author}{\bibfnamefont{K.}~\bibnamefont{W\'odkiewicz}},
  \bibinfo{journal}{Phys. Rev. A} \textbf{\bibinfo{volume}{41}},
  \bibinfo{pages}{369} (\bibinfo{year}{1990}).

\bibitem[{\citenamefont{Law et~al.}(2000)\citenamefont{Law, Chen, and
  Leung}}]{JC-pure-CK}
\bibinfo{author}{\bibfnamefont{C.~K.} \bibnamefont{Law}},
  \bibinfo{author}{\bibfnamefont{T.~W.} \bibnamefont{Chen}}, \bibnamefont{and}
  \bibinfo{author}{\bibfnamefont{P.~T.} \bibnamefont{Leung}},
  \bibinfo{journal}{Phys. Rev. A} \textbf{\bibinfo{volume}{61}},
  \bibinfo{pages}{023808} (\bibinfo{year}{2000}).

\bibitem[{\citenamefont{Lang et~al.}(1973)\citenamefont{Lang, Scully, and
  Lamb}}]{Lang_PRA}
\bibinfo{author}{\bibfnamefont{R.}~\bibnamefont{Lang}},
  \bibinfo{author}{\bibfnamefont{M.~O.} \bibnamefont{Scully}},
  \bibnamefont{and} \bibinfo{author}{\bibfnamefont{W.~E.} \bibnamefont{Lamb}},
  \bibinfo{journal}{Phys. Rev. A} \textbf{\bibinfo{volume}{7}},
  \bibinfo{pages}{1788} (\bibinfo{year}{1973}).

\bibitem[{\citenamefont{van~der Plank and Suttorp}(1996)}]{plank}
\bibinfo{author}{\bibfnamefont{R.~W.~F.} \bibnamefont{van~der Plank}}
  \bibnamefont{and} \bibinfo{author}{\bibfnamefont{L.~G.}
  \bibnamefont{Suttorp}}, \bibinfo{journal}{Phys. Rev. A}
  \textbf{\bibinfo{volume}{53}}, \bibinfo{pages}{1791} (\bibinfo{year}{1996}).

\bibitem[{\citenamefont{Ching et~al.}(1998)\citenamefont{Ching, Leung, van~den
  Brink, Suen, Tong, and Young}}]{QNM-RMP}
\bibinfo{author}{\bibfnamefont{E.~S.~C.} \bibnamefont{Ching}},
  \bibinfo{author}{\bibfnamefont{P.~T.} \bibnamefont{Leung}},
  \bibinfo{author}{\bibfnamefont{A.~M.} \bibnamefont{van~den Brink}},
  \bibinfo{author}{\bibfnamefont{W.~M.} \bibnamefont{Suen}},
  \bibinfo{author}{\bibfnamefont{S.~S.} \bibnamefont{Tong}}, \bibnamefont{and}
  \bibinfo{author}{\bibfnamefont{K.}~\bibnamefont{Young}},
  \bibinfo{journal}{Rev. Mod. Phys.} \textbf{\bibinfo{volume}{70}},
  \bibinfo{pages}{1545} (\bibinfo{year}{1998}).

\bibitem[{\citenamefont{Cohen-Tannoudji
  et~al.}(1992)\citenamefont{Cohen-Tannoudji, Dupont-Roc, and
  Grynberg}}]{Cohen}
\bibinfo{author}{\bibfnamefont{C.}~\bibnamefont{Cohen-Tannoudji}},
  \bibinfo{author}{\bibfnamefont{J.}~\bibnamefont{Dupont-Roc}},
  \bibnamefont{and} \bibinfo{author}{\bibfnamefont{G.}~\bibnamefont{Grynberg}},
  \emph{\bibinfo{title}{Atom-photon interactions}}, A Wiley-Interscience
  publication (\bibinfo{publisher}{Wiley}, \bibinfo{address}{New York},
  \bibinfo{year}{1992}).

\bibitem[{Res()}]{Resolvent-continuation}
\bibinfo{note}{Strictly speaking, Eqs.~(\ref{GEE2}) and (\ref{Gkk'}) are
  defined only in the upper half plane, and the analytic continuations of the
  functions in the second Riemann sheet are taken in the contour integration.
  In addition, the long time correction due to the branch point is also
  neglected.}

\bibitem[{\citenamefont{Turchette et~al.}(1995)\citenamefont{Turchette, Hood,
  Lange, Mabuchi, and Kimble}}]{Kimble-logic-gate}
\bibinfo{author}{\bibfnamefont{Q.~A.} \bibnamefont{Turchette}},
  \bibinfo{author}{\bibfnamefont{C.~J.} \bibnamefont{Hood}},
  \bibinfo{author}{\bibfnamefont{W.}~\bibnamefont{Lange}},
  \bibinfo{author}{\bibfnamefont{H.}~\bibnamefont{Mabuchi}}, \bibnamefont{and}
  \bibinfo{author}{\bibfnamefont{H.~J.} \bibnamefont{Kimble}},
  \bibinfo{journal}{Phys. Rev. Lett.} \textbf{\bibinfo{volume}{75}},
  \bibinfo{pages}{4710} (\bibinfo{year}{1995}).

\bibitem[{pro()}]{prob}
\bibinfo{note}{In Ref.~\cite{Kimble-logic-gate}, the values of $\kappa$ and
  $\gamma$ refer to the decay rate of the corresponding amplitudes. However, in
  this paper, we use $\kappa$ and $\gamma$ to denote the decay rates of
  probabilities, and hence a factor of $2$ is multiplied. In addition, while
  the atoms in the original setup interact only with one polarization mode of
  the photon, for a qualitative estimation, we will let
  $\lambda_{L(R)}/2\pi\sim 20{\rm MHz}$ whenever convenient.}

\bibitem[{\citenamefont{Brattke et~al.}(2003)\citenamefont{Brattke, Guth{\"
  o}hrlein, Keller, Lange, Varcoe, and Walther}}]{Brattke}
\bibinfo{author}{\bibfnamefont{S.}~\bibnamefont{Brattke}},
  \bibinfo{author}{\bibfnamefont{G.~R.} \bibnamefont{Guth{\" o}hrlein}},
  \bibinfo{author}{\bibfnamefont{M.}~\bibnamefont{Keller}},
  \bibinfo{author}{\bibfnamefont{W.}~\bibnamefont{Lange}},
  \bibinfo{author}{\bibfnamefont{B.}~\bibnamefont{Varcoe}}, \bibnamefont{and}
  \bibinfo{author}{\bibfnamefont{H.}~\bibnamefont{Walther}},
  \bibinfo{journal}{J. Mod. Opt.} \textbf{\bibinfo{volume}{50}},
  \bibinfo{pages}{1103} (\bibinfo{year}{2003}).

\end{thebibliography}

\newpage

\end{document}